\documentclass[prl,aps,twocolumn]{revtex4-1}
\usepackage{amsmath}
\usepackage{amssymb}
\usepackage{graphicx}
\usepackage{bbold}

\newcommand{\beqa}{\begin{eqnarray}}
\newcommand{\eeqa}{\end{eqnarray}}

\begin{document}
\title{$\mathcal{PT}$ spectroscopy of the Rabi problem} 
\author{Yogesh N. Joglekar$^1$, Rahul Marathe$^2$, P. Durganandini$^2$, and Rajeev K. Pathak$^2$} 
\affiliation{$^1$Department of Physics, Indiana University Purdue University Indianapolis (IUPUI), 
Indianapolis, Indiana 46202, USA\\ 
$^2$Department of Physics, University of Pune, Ganeshkhind, Pune 411007, India}

\date{\today}
\begin{abstract}
We investigate the effects of a time-periodic, non-hermitian, $\mathcal{PT}$-symmetric perturbation on a system with two (or few) levels, and obtain its phase diagram as a function of the perturbation strength and frequency.  We demonstrate that when the perturbation frequency is close to one of the system resonances, even a vanishingly small perturbation leads to $\mathcal{PT}$ symmetry breaking. We also find a restored $\mathcal{PT}$-symmetric phase at high frequencies, and at moderate perturbation strengths, we find multiple frequency windows where $\mathcal{PT}$-symmetry is broken and restored. Our results imply that the $\mathcal{PT}$-symmetric Rabi problem shows surprisingly rich phenomena absent in its hermitian or static counterparts.  
\end{abstract}
\maketitle

\noindent{\it Introduction.} A two-level system coupled to a sinusoidally varying potential is a prototypical example of a time-dependent, exactly solvable Hamiltonian, with profound implications to atomic, molecular, and optical physics~\cite{sakurai}. When the frequency of perturbation $\omega$ is close to the characteristic frequency $\Delta$ of the two-level system - near resonance - the system undergoes complete population inversion for an arbitrarily small strength $\gamma$ of the potential~\cite{rabi}. The implications of this result to spin magnetic resonance, Rabi flopping~\cite{lasers}, and its generalization, namely the Jaynes-Cummings model~\cite{jc,jc2}, have been extensively studied over the past half century~\cite{reviewjc1,reviewjc2}. Surprisingly, the quantum Rabi problem, where the full quantum nature of the perturbing bosonic field is taken into account, has only been recently solved~\cite{braak}. 

The two-level model is useful because it is applicable to many-level systems when the perturbation frequency is close to or resonant with a {\it single pair of levels}. As the detuning away from resonance $|\Delta-\omega|$ increases, the perturbation strength necessary for population inversion increases linearly with it; in a many-level system, with increased potential strength, transitions to other levels have to be taken into account and the resultant problem is not exactly solvable. Therefore, understanding the behavior of a system in the entire parameter space $(\gamma,\omega)$ requires analytical and numerical approaches. All of these studies are restricted to hermitian potentials. 

In recent years, discrete Hamiltonians with a hermitian tunneling term $H_0$ and a non-hermitian perturbation $V$ that are invariant under combined parity and time-reversal ($\mathcal{PT}$) operations have been extensively investigated~\cite{znojil1,znojil2,bendix,song,mark,avadh,derek,longhi1}. The spectrum $\epsilon_\lambda$ of a $\mathcal{PT}$-symmetric Hamiltonian is real when the strength $\gamma$ of the non-hermitian perturbation is smaller than a threshold $\gamma_{PT}$ set by the hermitian tunneling term. Traditionally, the emergence of complex-conjugate eigenvalues that occurs when the threshold is exceeded, $\gamma>\gamma_{PT}$, is called $\mathcal{PT}$ symmetry breaking~\cite{bender1,bender2,bender3}. It is now clear that $\mathcal{PT}$-symmetric Hamiltonians represent open systems with balanced gain and loss, and $\mathcal{PT}$ symmetry breaking is a transition from a quasiequilibrium state ($\mathcal{PT}$-symmetric state) to a state with broken reciprocity ($\mathcal{PT}$-broken state)~\cite{review}. Recent experiments on optical waveguides~\cite{expt1,expt2,expt3} and resonators~\cite{expt4} with amplification and absorption have shown that $\mathcal{PT}$ systems display a wealth of novel phenomena~\cite{uni1,uni2} that are absent in closed or purely dissipative systems. We note that all experiments and most of the theoretical work, with few exceptions~\cite{moiseyev1,yuri,longhitime}, have only explored systems with static gain and loss potentials.  

And what of a system perturbed by a time-periodic gain-loss potential $V(t)$? What is the criterion  for $\mathcal{PT}$-symmetry breaking? What is the analog of Rabi flopping in such a case? We answer these questions by investigating small, $N$-site lattices perturbed by a pair of balanced gain-loss potentials $\pm i\gamma\cos(\omega t)$ located at parity symmetric sites. Such systems can be realized in coupled waveguides with a complex refractive index~\cite{expt2,expt3,yuri} that varies along the propagation direction, or in coupled resonators~\cite{expt4}.  

Our primary results are as follows: i) The system can be in ``$\mathcal{PT}$-broken phase'' even if the spectrum $\epsilon_\lambda(t)$ of the Hamiltonian $H(t)$ is real at all times. ii) Near every resonance, the $\mathcal{PT}$ symmetric threshold is reduced from its static value $\gamma_{PT}$ to the detuning, $\gamma_{PT}(\omega)\propto |\omega-\Delta|$; in particular, it vanishes at the resonance. iii) For any gain-loss strength including $\gamma\gg\gamma_{PT}$, the $\mathcal{PT}$-symmetric phase is restored at high frequencies $\omega>\omega_c\propto\gamma$. iv) At intermediate strengths, $\gamma\sim\gamma_{PT}$, the $\mathcal{PT}$ symmetry is broken in multiple windows in the frequency domain. Thus, a harmonic, $\mathcal{PT}$-symmetric perturbation provides a new spectroscopic tool for investigating the level structure of a system. 

\begin{figure*}[tbhp]
\centering
\includegraphics[width=1.0\columnwidth]{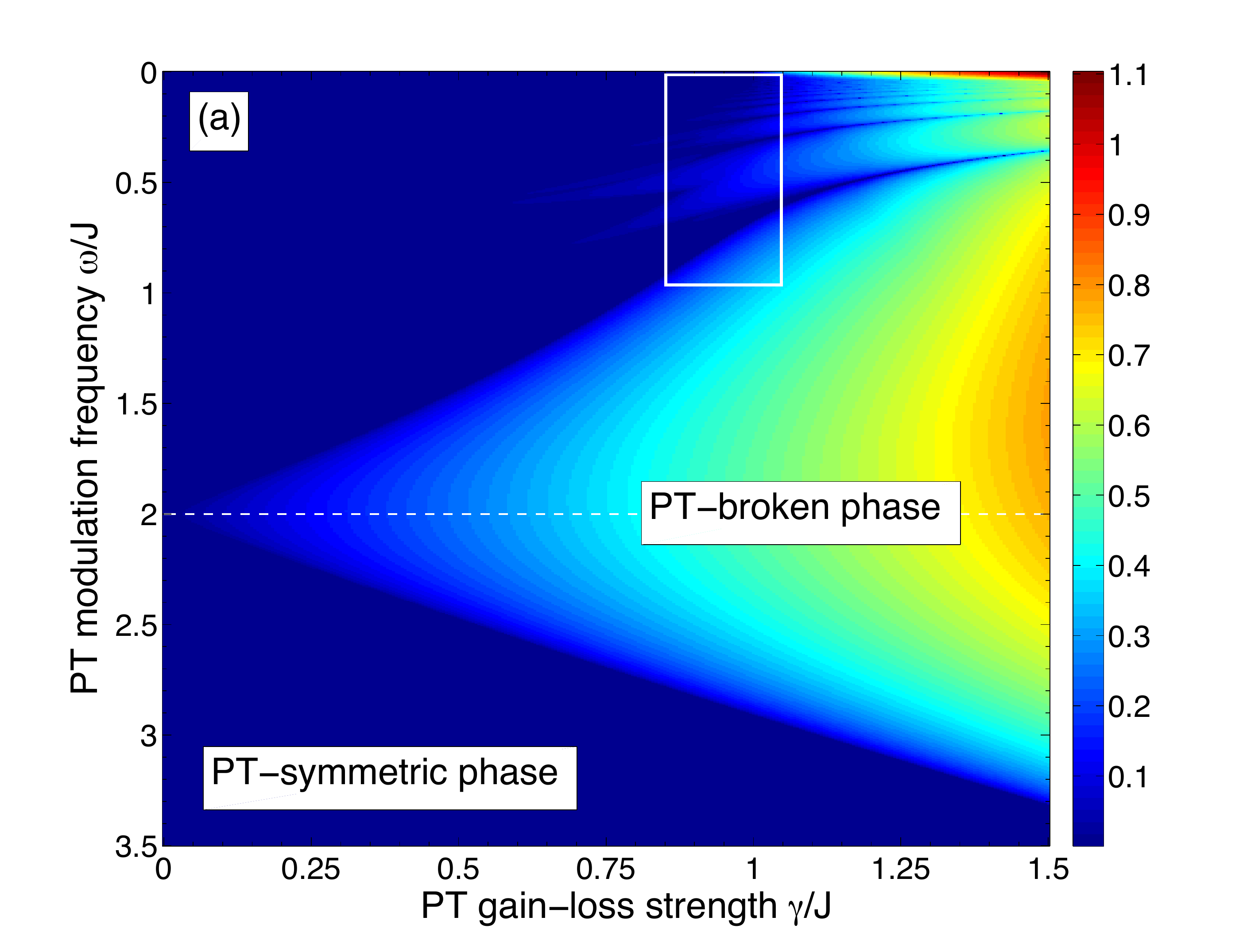}
\includegraphics[width=1.0\columnwidth]{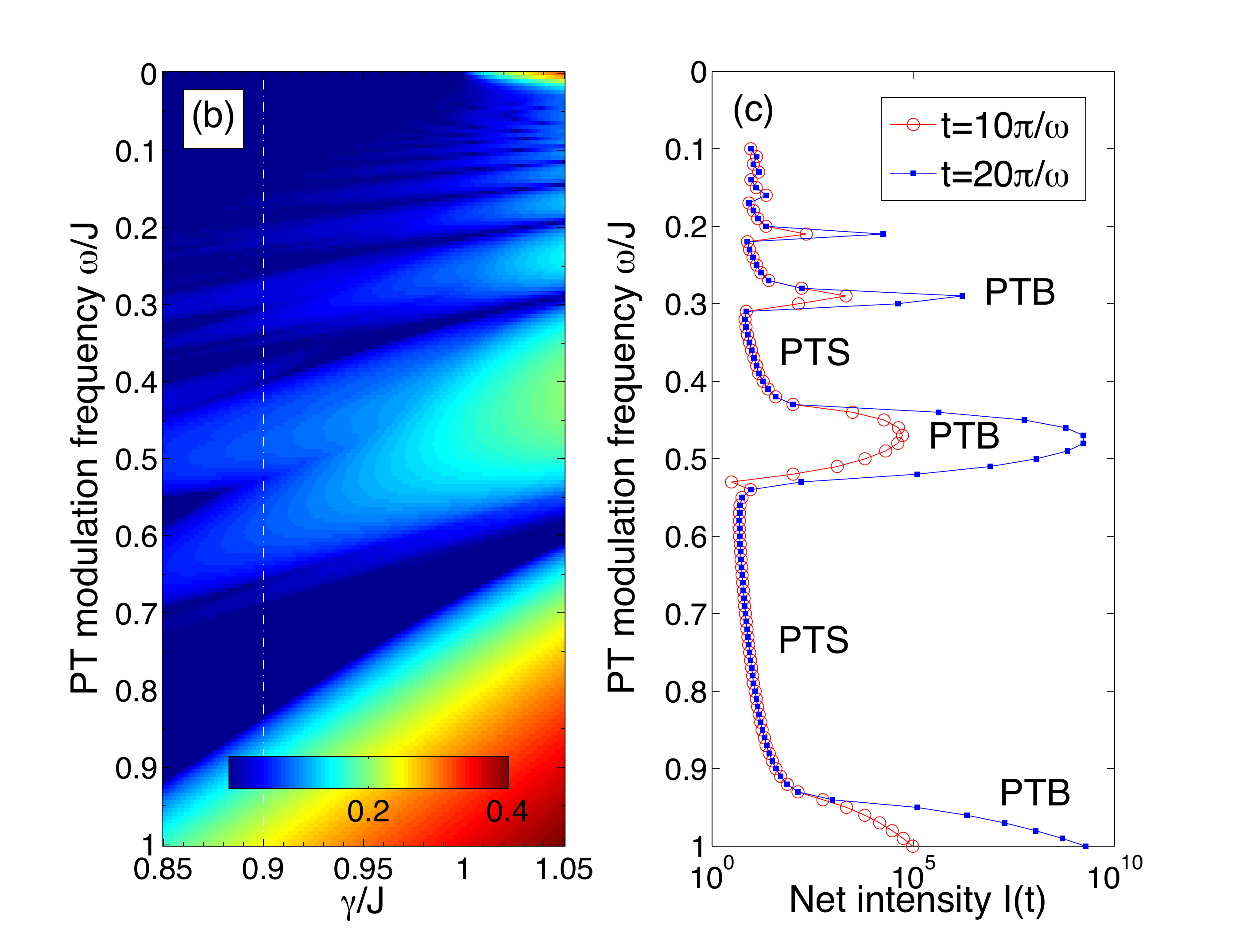}
\caption{(color online) (a) $\mathcal{PT}$ phase diagram of a two-level system in the $(\gamma,\omega)$ plane; plotted is the largest imaginary part of the spectrum of $\mathcal{H}$. The static threshold $\gamma_{PT}/J=1$ is suppressed down to zero when the perturbation frequency matches a resonance, $\omega/J=2$. (b) A close-up of the area marked by the white rectangle in panel (a) shows that for moderate $\gamma/J\sim 1$, multiple $\mathcal{PT}$-symmetric (PTS) and $\mathcal{PT}$-broken (PTB) regions occur when $\omega$ is varied. (c) Maximum of the net intensity $I(t)=\langle\psi(0)| G^\dagger(t) G(t)|\psi(0)\rangle$ at two cutoff times $t=10\pi/\omega$ (open red circles) and $t=20\pi/\omega$ (solid blue squares), on horizontal logarithmic axis, as a function of $\omega$ on the vertical axis, at $\gamma/J=0.9$ (white dot-dashed line in panel (b)). PTB regions are distinguished by a clear dependence of $I_{\max}$ on the cutoff.}
\label{fig:n2}
\vspace{-3mm}
\end{figure*}
\noindent{\it $\mathcal{PT}$ phase diagram.} The Hamiltonian for an $N$-site lattice with constant tunneling is 
\begin{equation}
\label{eq:h0}
H_0=-\hbar J\sum_{x=1}^{N-1} \left(|x\rangle\langle x+1|+|x+1\rangle\langle x|\right),
\end{equation}
 where $|x\rangle$ is a normalized state localized on site $x$, $J$ is the tunneling rate, and $\hbar=h/(2\pi)$ is the scaled Planck's constant. The action of the parity operator on the lattice is given by $x\rightarrow \bar{x}=N+1-x$ and the antilinear time-reversal operator acts as $i\rightarrow -i$. 
The spectrum of $H_0$ is given by $\epsilon_n=-2\hbar J\cos(k_n)=-\epsilon_{\bar{n}}$ and the normalized eigenfunctions are $\psi_n(x)=\langle x|n\rangle=\sin(k_n x)/\sqrt{1+1/N}$ where $k_n=n\pi/(N+1)$ with $1\leq n\leq N$. The energy differences $\hbar\Delta_{nm}=\epsilon_n-\epsilon_m>0$ define the possible resonances for this $N$-level system. Motivated by the Rabi problem, here we will only consider $N=\{2,3,4\}$. This system is perturbed by a balanced gain-loss potential 
\begin{equation}
\label{eq:v}
V(t)=i\hbar\gamma\cos(\omega t)\left(|x_0\rangle\langle x_0| - |\bar{x}_0\rangle\langle \bar{x}_0|\right)\neq V^\dagger(t). 
\end{equation}
Eq.(\ref{eq:v}) implies that at time $t=0$, $x_0$ is the gain or amplification site and $\bar{x}_0$ is the loss or absorption site. The non-hermitian potential satisfies $\mathcal{PT}V(t)\mathcal{PT}=V(t)$. The total Hamiltonian $H(t)=H_0+V(t)$ is periodic in time, i.e. $H(t+2\pi/\omega)=H(t)$, and its properties are best analyzed via its Floquet counterpart, $\mathcal{H}=-i\hbar\partial_t+H$~\cite{floquet,shirley,moiseyev2}. In the frequency domain, the non-Hermitian, $\mathcal{PT}$-symmetric Floquet Hamiltonian is given by 
\begin{eqnarray}
\label{eq:hf}
\mathcal{H}^{p,q}_{x,x'} & = & -ip\hbar\omega\delta_{p,q}\delta_{x,x'}-\delta_{p,q}\hbar J(\delta_{x,x'+1}+\delta_{x,x'-1}) \nonumber\\
& + & i\hbar\gamma\delta_{x,x'}(\delta_{p,q+1}+\delta_{p,q-1})(\delta_{x,x_0}-\delta_{x,\bar{x}_0}) 
\end{eqnarray}
where $p,q\in\mathbb{Z}$ denote the Floquet band indices. In practice, a truncated Floquet Hamiltonian with $|p|\leq N_f$ is used, so that $\mathcal{H}$ is an $N(2N_f+1)\times N(2N_f+1)$ matrix.

We define $\mathcal{PT}$-symmetry breaking as {\it the emergence of complex-conjugate eigenvalues for the Floquet Hamiltonian} $\mathcal{H}$~\cite{moiseyev1}. As we will show below, this can occur even if the instantaneous eigenvalues $\epsilon_\lambda(t)$ of the time-dependent Hamiltonian $H(t)$ are purely real over the entire period $T_\omega=2\pi/\omega$. 

Figure~\ref{fig:n2} is the $\mathcal{PT}$-symmetric phase diagram of a two-level system in the $(\gamma,\omega)$ plane. Figure~\ref{fig:n2}(a) shows the maximum imaginary part of the spectrum of the Hamiltonian $\mathcal{H}$ with 101 Floquet bands. The region with zero imaginary part (dark blue) is the $\mathcal{PT}$-symmetric phase and the region with nonzero imaginary part (all other colors) is the $\mathcal{PT}$-broken phase;  the static, $\omega=0$, threshold is given by $\gamma_{PT}/J=1$. Figure.~\ref{fig:n2}(a) has three universal features that appear in systems with larger $N$ as well. The first is a vanishingly small $\mathcal{PT}$-symmetric threshold $\gamma_{PT}(\omega)$ that occurs when $\omega$ is close to the single resonance frequency for the system, $\Delta_{21}=2J$. The second is the the emergence of $\mathcal{PT}$-symmetric phase that occurs at large frequencies~\cite{longhitime} $\omega>\omega_c\propto\gamma$ for any gain-loss strength including $\gamma/J\geq 1$. The third feature is the presence of multiple windows along the frequency axis such that the $\mathcal{PT}$-symmetry is broken within each window. Fig.~\ref{fig:n2}(b) is a higher resolution close-up of the parameter space marked by the white rectangle in Fig.~\ref{fig:n2}(a). It shows that for a fixed $\gamma$, as the frequency of the $\mathcal{PT}$ potential is changed, multiple $\mathcal{PT}$-symmetric and $\mathcal{PT}$-broken regions emerge. {\it These regions are present both below and above the static threshold}.

A complementary method to distinguish the $\mathcal{PT}$-broken (PTB) region from the $\mathcal{PT}$-symmetric (PTS) region is to track the net intensity $I(t)=\langle \psi(t)|\psi(t)\rangle=\langle\psi(0)| G^\dagger(t) G(t)|\psi(0)\rangle$ of an initially normalized state $|\psi(0)\rangle$. We obtain the non-unitary time-evolution operator 
\begin{equation}
\label{eq:g}
G(t)=Te^{-{\frac{i}{\hbar}\int_0^t dt' H(t')}}\approx
\prod_{k=1}^{t/\delta t} e^{-\frac{i}{\hbar}\delta t H(k\delta t)}, 
\end{equation}
where the discretization time-step $\delta t/T_\omega\ll 1$ is chosen to ensure that $G(t)$ is independent of it. Figure~\ref{fig:n2}(c) shows the maximum intensity $I_{\max}$ reached before time $t$ at cutoff times $t=5T_\omega$ (open red circles) and $t=10T_\omega$ (solid blue squares), as a function of the $\mathcal{PT}$-perturbation frequency $\omega$ on the vertical axis. These results are for $\gamma/J=0.9$, initial state localized on the first site, and a time-step $\delta t/T_\omega=10^{-5}$. In the $\mathcal{PT}$-symmetric region, $I(t)$ undergoes bounded oscillations and therefore $I_{\max}$ is the same for the two time-cutoffs. In a sharp contrast, in the $\mathcal{PT}$-broken region, $I(t)$ increases exponentially with time and $I_{\max}$ doubles on the logarithmic scale when the cutoff is doubled. Thus, both approaches show the existence of multiple frequency windows where $\mathcal{PT}$ symmetry is broken; note that the phase-boundaries in Fig.~\ref{fig:n2}(c) do not {\it exactly} match those in Fig.~\ref{fig:n2}(b) due to the finite time-cutoff. 

We remind the reader that when $\gamma/J\leq 1$, the $2\times 2$ Hamiltonian $H(t)=i\hbar\gamma\cos(\omega t)\sigma_z -\hbar J\sigma_x$ has a purely real spectrum $\epsilon_\lambda(t)=\pm\hbar[J^2-\gamma^2\cos^2(\omega t)]^{1/2}$ at all times, and yet, the norm of the  time-evolution operator $G(t)$ is either bounded or exponential-in-time at different frequencies~\cite{moiseyev1}. 

\begin{figure}[thpb]
\centering
\includegraphics[angle=0,width=\columnwidth]{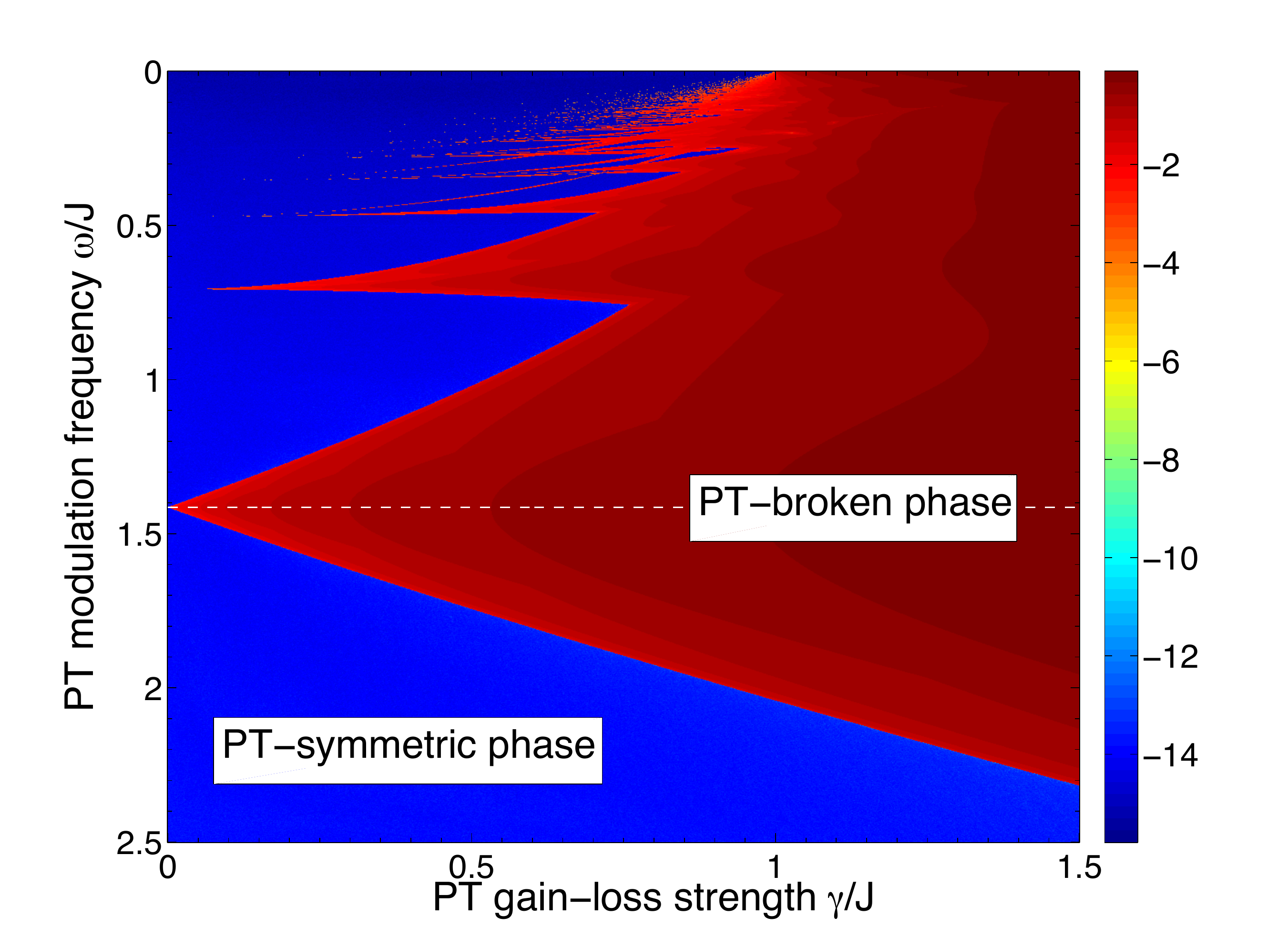}
\caption{(color online) $\mathcal{PT}$-symmetric phase diagram of a three-level system; shown is the base-10 log of the maximum imaginary part of the spectrum of $\mathcal{H}$. Its three features - a linearly vanishing $\mathcal{PT}$-threshold at the resonance $\omega/J=\sqrt{2}\approx 1.41$, a restored $\mathcal{PT}$-symmetric phase at high frequencies, and multiple frequency windows with $\mathcal{PT}$-symmetric and $\mathcal{PT}$-broken phases at moderate $\gamma/J$ - are universal.} 
\label{fig:n3}
\end{figure}
Figure~\ref{fig:n3} shows the $\mathcal{PT}$ phase diagram of a three-level system obtained by using 81 Floquet bands. We plot the base-10 logarithm of the largest imaginary part of eigenvalues of $\mathcal{H}$ to easily distinguish the $\mathcal{PT}$-symmetric phase (blue) and $\mathcal{PT}$-broken phase (red). For a three-level system, the unperturbed spectrum is given by $\epsilon_n=\{0,\pm\sqrt{2}J\}$, and has two resonance frequencies $\Delta_{21}=\sqrt{2}J=\Delta_{32}$ and $\Delta_{31}=2\sqrt{2}J$. Fig.~\ref{fig:n3} shows a linearly vanishing $\gamma_{PT}(\omega)$ at the first resonance $\omega/J=\sqrt{2}$, a $\mathcal{PT}$-restored phase at high frequencies, and a number of frequency windows where $\mathcal{PT}$ symmetry is broken at moderate values of $\gamma/J\lesssim 1$. There is no $\mathcal{PT}$-broken region near the second resonance, $\omega/J=2\sqrt{2}$ (not shown) because $V(t)$ does not connect states with same parity. Thus, the phase diagram for a three-level system shares the fundamental characteristics of that for a two-level system. 

\begin{figure}[thpb]
\centering
\includegraphics[angle=0,width=\columnwidth]{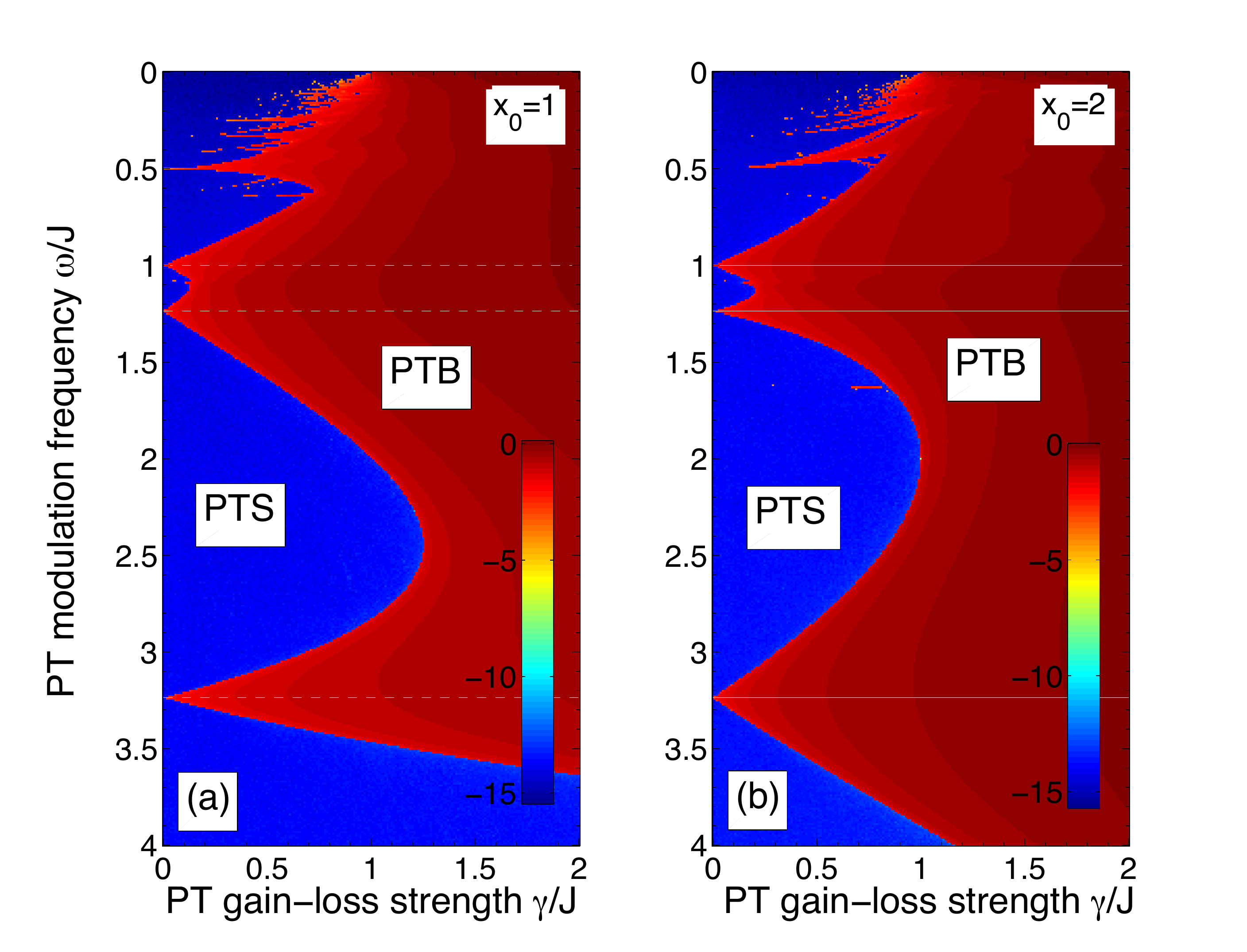}
\caption{(color online) Phase diagram of a 4-site lattice with gain-loss potential at the ends, $x_0=1$ (a), or in the center, $x_0=2$ (b). Both panels show a restored $\mathcal{PT}$-symmetric phase at high $\omega$, and multiple frequency windows with PTS and PTB regions at moderate $\gamma/J$. They also show a linearly vanishing threshold $\gamma_{PT}(\omega)$ at each of the three relevant resonances, marked by dashed (a) or solid (b) white lines.} 
\label{fig:n4}
\end{figure}
Lastly, we consider a four-level system, $N=4$. In this case, the resonances between states with opposite parity occur at $\Delta_{21}/J=1=\Delta_{34}/J$, $\Delta_{23}/J=(\sqrt{5}-1)\approx 1.236$, and $\Delta_{14}/J=(\sqrt{5}+1)\approx 3.236$. There are two inequivalent locations for the gain-loss potential, $x_0=1$ and $x_0=2$, and both have the same static threshold $\gamma_{PT}/J=1$. In the first case, only center-two of the four eigenvalues become complex at the threshold, whereas in the second case all four simultaneously become complex~\cite{jake}. Figure~\ref{fig:n4} shows the $\mathcal{PT}$ phase diagram obtained with 41 Floquet bands. Both panels, (a) and (b), demonstrate the three salient features discussed earlier for both two and three-level systems.   


\noindent{\it Understanding the phase boundaries.} We now derive the $\mathcal{PT}$-phase boundaries at occur at high frequencies or vanishingly small $\gamma$ near a resonance. The natural time-scale for an unperturbed system is proportional to $1/J$ and its static $\mathcal{PT}$ breaking threshold is also set by $J$. At high frequencies $\omega/J\gg 1$, the rapidly varying potential $i\gamma\cos(\omega t)$ is replaced by its average over the characteristic time-scale, $\gamma_{\mathrm{av}}\propto (\gamma/\omega) J$. For any gain-loss strength $\gamma$, no matter how large, increasing the frequency reduces the effective strength $\gamma_\mathrm{av}$ and thus restores the $\mathcal{PT}$-symmetric phase.  The slope of the linear phase-boundary in the region $\omega/J\gg 1$ will depend upon the number of levels $N$ and the location $x_0$ of the gain-loss potential, but the linear behavior of the phase boundary is universal. 

Next we derive the cone-shaped phase boundary that occurs at small $\gamma$ in the neighborhood of a resonance $\omega\sim\Delta_{nm}$. For a state $|\psi(t)\rangle = \sum_n c_n(t) \exp(-i\epsilon_n t/\hbar)|n\rangle$, the interaction-picture equation of motion for the level-occupation coefficients $c_n(t)$ is given by 
\begin{eqnarray}
\label{eq:rabi1}
i\partial_t c_n(t) & = & \sum_{m=1}^N V_{nm}(t) e^{+i\Delta_{nm}t}c_m(t),\\
\label{eq:rabi2}
V_{nm}(t) & =  &i \gamma\cos(\omega t)[1-(-1)^{n+m}]\langle n|x_0\rangle\langle x_0|m\rangle.
\end{eqnarray}

For a two-level system, when $\omega\approx \Delta_{21}=2J$, averaging over high-frequency terms simplifies Eq.(\ref{eq:rabi1}) to 
\begin{equation}
\label{eq:final}
\partial_t^2c_{1,2}(t)+\left[(\omega/2-J)^2-(\gamma/2)^2\right]c_{1,2}(t)=0.
\end{equation}
Eq.(\ref{eq:final}) implies that when $|\omega-2J|>\gamma$, the coefficients $c_1(t)$ and $c_2(t)$ oscillate in time and remain bounded, and the system is in the $\mathcal{PT}$-symmetric phase. When $|\omega-2J|<\gamma$, $c_{1,2}(t)$ increase with time exponentially, and the system is in the $\mathcal{PT}$-broken phase. Thus, $\mathcal{PT}$-symmetric phase boundary is given by $\gamma_{PT}(\omega)=|\omega-2J|$, and the threshold gain-loss strength vanishes as $\omega\rightarrow \Delta_{21}=2J$. A visual inspection of Fig.~\ref{fig:n2}(a) shows that, indeed, the slope of the phase-boundary lines fanning away from $\omega=2J$ is one. Eq.(\ref{eq:final}) also implies that along the cone-shaped phase boundary, the net intensity $I(t)$ grows quadratically with time~\cite{expt1,expt4,review}. 

When $N=3$ a similar analysis with three equidistant levels implies that near resonance $c_2(t)$ satisfies a third-order differential equation,  $\partial^3_t c_2(t)+\left[(\omega-\sqrt{2}J)^2-(3\gamma/4)^2\right]\partial_t c_2(t)=0$. Therefore, the phase-boundary separating the $\mathcal{PT}$-broken region from the $\mathcal{PT}$-symmetric region is given by $\gamma_{PT}(\omega)=(4/3)|\omega-\sqrt{2}J|$. This, too, can be verified by a visual inspection of Fig.~\ref{fig:n3}. Our analysis also predicts that along this phase boundary, the net intensity of an initially normalized state increases quartically with time, i.e. $I(t)\propto t^4$, because $\partial^3_t c_2(t)=0$. In an $N$-level system, our analysis is applicable to a pair of levels $m,n$ when the $\mathcal{PT}$-perturbation frequency is close to the resonance that connects those levels, $\omega\sim\Delta_{nm}$. 

Thus, {\it a vanishingly small $\mathcal{PT}$ perturbation induces $\mathcal{PT}$-symmetry breaking when the frequency of the perturbation is close to a resonance}. Near resonance, the spatial oscillations of a state match the gain-loss temporal oscillations; as a result, it spends most of the time on the gain-medium site, leading to an exponential growth in the net intensity. 


\noindent{\it Conclusion.} In this paper, we have proposed the $\mathcal{PT}$-symmetric Rabi model. We have shown that a harmonic, gain-and-loss perturbation leads to a rich $\mathcal{PT}$ phase diagram with three salient features. Among them is the existence of multiple frequency windows 
in which $\mathcal{PT}$-symmetry is broken and restored. Time-dependent $\mathcal{PT}$ potentials have been extensively investigated in continuum one-dimensional optical structures~\cite{uni1,uni2}. Our results show that such potentials are a surprising spectroscopic probe, where the phase of the system - $\mathcal{PT}$-broken  or $\mathcal{PT}$-symmetric - denotes the proximity of the perturbation frequency to a resonance of the system. 

Although we have focused only on few-level systems here, our results are applicable to larger lattices, particularly in the vicinity of a resonance. Deep in the $\mathcal{PT}$-broken phase, at long times, nonlinear effects also become relevant, although they do not affect our findings. 


This work was supported by NSF DMR-1054020 (YJ), Department of Science and Technology, India (RM), BUCD University of Pune (PD), and University of Pune research grant BCUD/RG/9 (RP). YJ thanks the hospitality of University of Pune where this work began. 

\end{document}